\documentclass[a4paper]{ISOStyle}
\usepackage{epsfig}

\begin{document}

\title{The Mid-Infrared Spectra of Interacting Galaxies: From ISO to SIRTF}

\author{V. Charmandaris\inst{1} 
	\and O.Laurent\inst{2} 
	\and I.F. Mirabel\inst{3} 
	\and P. Gallais\inst{3} 
	\and J. Houck\inst{1}} 
\institute{Cornell University, IRS Science Center, Ithaca NY 14853, USA
  	\and Max-Planck-Institut f\"ur Extraterrestrische Physik, Postfach 1603, 85740 Garching,  Germany
	\and CEA/DSM/DAPNIA, Service d'Astrophysique, F-91191 Gif-Sur-Yvette, France}

\maketitle 

\begin{abstract}

We present mid-infrared (5--16$\mu m$) images and spectra of a
sequence of interacting galaxies, observed by ISOCAM. The galaxies
were selected as being at progressive stages in the time evolution of
a merging event, following what is known as Toomre's ``merger
sequence'', and having no detected contribution from an active
galactic nucleus (AGN) in their mid-infrared spectrum. To trace the
intensity of the global star formation in those galaxies, we use the
ratio of the 15$\mu m$ to 7$\mu m$ flux. Our analysis indicates that
this ratio increases from $\sim$ 1 to $\sim$ 5 as galaxies move from
the pre-starburst to the merging/starburst phase only to decrease to
$\sim$ 1 again in the post-starburst phase of the evolved merger
remnants. Moreover, we find that the variation of this ratio is well
correlated with the one of the IRAS 25$\mu m$/12$\mu m$ and 60$\mu
m$/100$\mu m$ flux ratios. The potential to test and improve upon
these results using the Infrared Spectrograph (IRS) on board SIRTF is
discussed.


\end{abstract}

\section{INTRODUCTION}

One of the major steps in the understanding of galaxy evolution was
the realization that tails and bridges are the result of galaxy
interactions (\cite{tt}).  It was also proposed by Toomre (1977) to
use the morphology of the observed tidal features and the separation
between the galaxies in order to create a ``merging sequence'' of 11
peculiar NGC galaxies, also found in the Arp atlas. Ever since,
improvements in numerical modeling of the stellar and gaseous
component in galaxies have clearly demonstrated that galaxy
interactions cause large scale instabilities in the galactic disks
leading to the formation of transient bars which drive the gas into
the center of the galaxies (\cite{bn}).  Furthermore, numerous
multi-wavelength studies of those systems (see Hibbard 1995; Schweizer
1998 and references therein) have been performed in effort to better
understand phenomena such as starburst and AGN activity, as well as
mass transfers and morphological transformations associated with
interacting galaxies. One of the major quests in those studies
remained the identification of observational characteristics which
could be used as alternatives of assigning an ``age'' to the event of
the interaction (i.e. \cite{ss92}). The discovery by IRAS of the class
of luminous IR galaxies which harbor of obscured massive starbursts
(\cite{soifer}) and the revelation later on that they are also
interacting/merging systems (\cite{sanders}), attracted further
attention to this problem (see \cite{sm} for a review).

In this paper we examine the global star formation activity in a
sample of interacting galaxies as it becomes evident in the
mid-infrared via the heating of the dust.

\section{THE SAMPLE}

The galaxies of our sample were part of the ISOCAM (\cite{cesarsky})
active galaxy proposal CAMACTIV (P.I. F. Mirabel). The galaxies were
observed in the spectro-imaging mode with the Circular Variable Filter
(CVF) or, for weaker sources, in the raster mode with broad band
filters. Information on the whole CAMACTIC sample as well as on the
observational techniques can be found in \cite*{thesis} and
\cite*{olivier}.  The standard data reduction procedures pertinent to
ISOCAM data were followed resulting in a photometric accuracy of 20\%.

The galaxies were selected with an {\it apriori knowledge} of their
stage of interaction and also based on the fact that the AGN
contribution is negligible in the mid-infrared (\cite{olivier}).  They form an
evolution sequence from galaxies in early stages of interaction:
NGC4676, NGC 3263, and NGC 520; to galaxies approaching a merger
stage: NGC3256, NGC6240 and Arp220; and finally to the classified late
merger remnants: NGC7252 and NGC3921. The selection process and
details on the galaxies are beyond the scope of this paper and will be
presented in \cite*{vassilis}.

\section{DISCUSSION}

\subsection{The ISO results}

\begin{figure*}[!ht]
\begin{center}
 \resizebox{0.8\hsize}{!}{\includegraphics{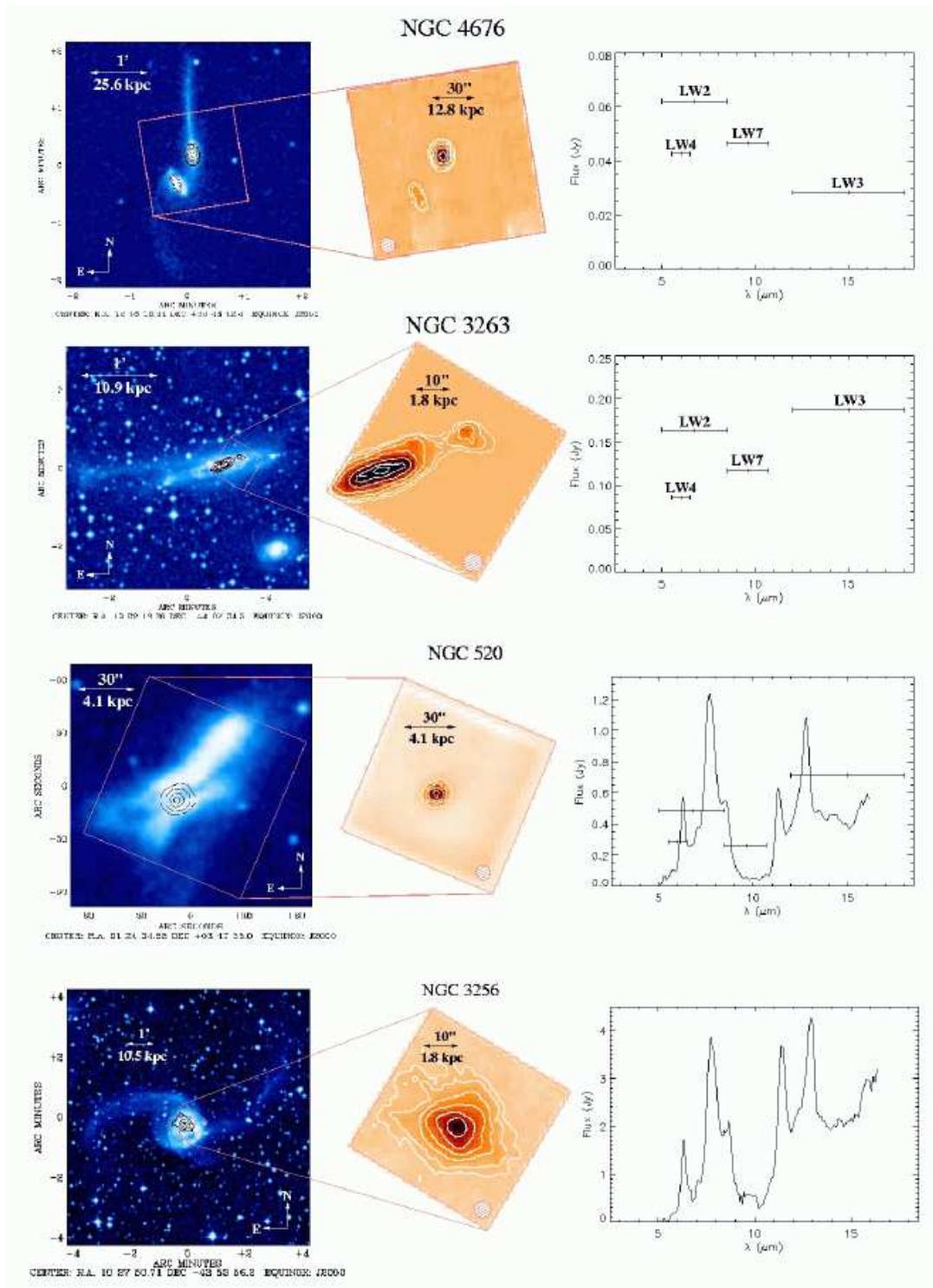}}
\end{center} 
  \caption{In this figure we present four of the eight galaxies of our
  sample found in increasing stages of interaction. From NGC\,4676 at
  the top to NGC\,3256 at the bottom. For each galaxy we include an
  optical DSS image on the left, marked with the box imaged by ISOCAM,
  a 7$\mu$m image in the middle and the mid-infrared spectrum of the
  galaxy on the right. Note how the flux beyond 10$\mu$m progressively
  increases comparing to the strength of the UIB features. The
  horizontal bars indicate the width of several of the broad band
  filters used in the observations. NGC\,4676 and NGC\,3632 as well as
  NGC\,3921 (see Fig. 2) were observed only in broad band filter mode.
\label{fig1}}
\end{figure*}

\begin{figure*}[!ht]
\begin{center}
 \resizebox{0.8\hsize}{!}{\includegraphics{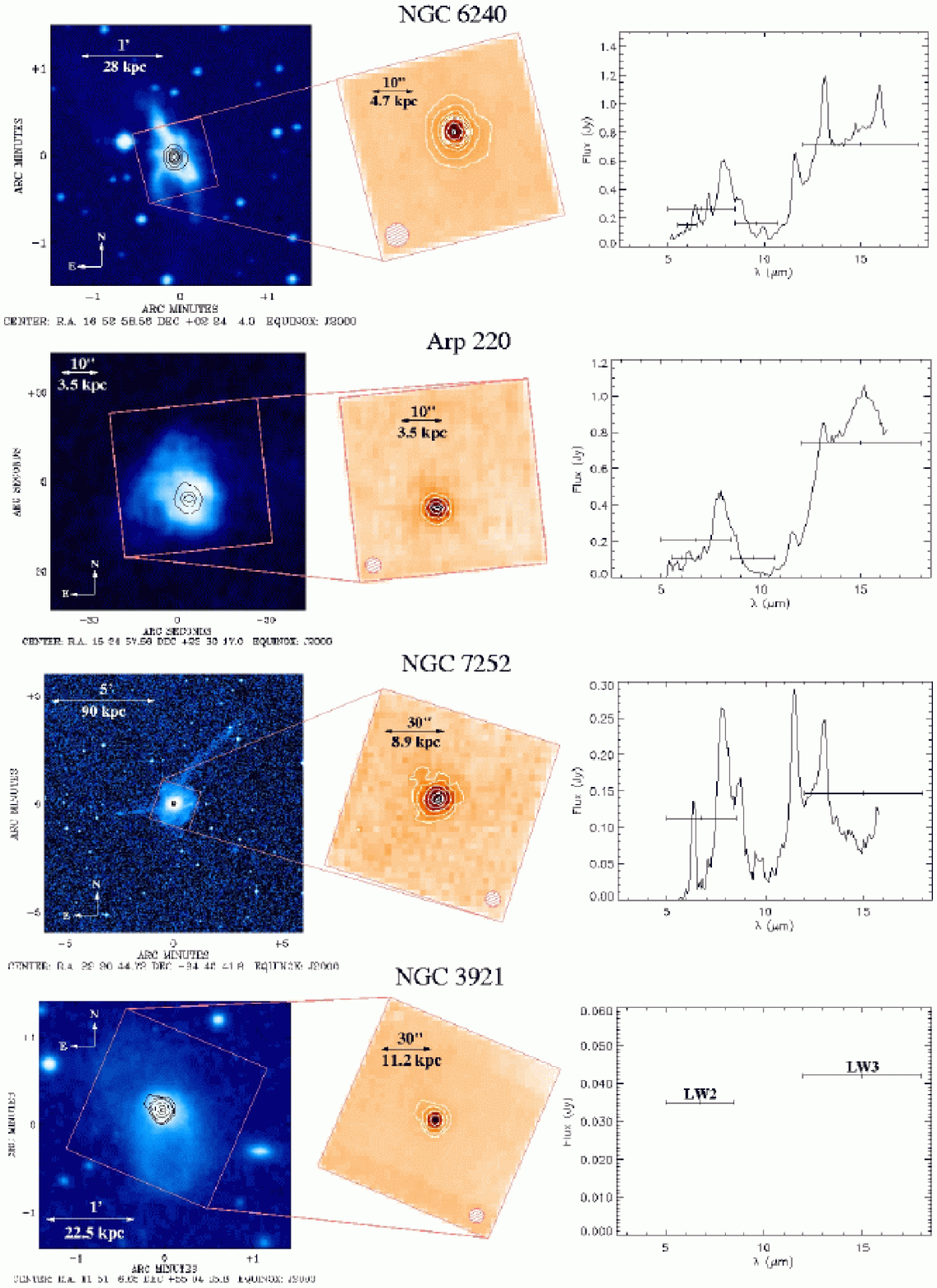}}
\end{center}
  \caption{ Same as in Fig. 1 for the remaining four galaxies. The
  mid-infrared continuum reaches its peak emission in Arp220 and
  progressively decreases in NGC7252 and NGC3921. The solid circle in
  the mid-infrared 7$\mu$m images indicates the FWHM of the point
  spread function.
\label{fig2}}
\end{figure*}

In Figures \ref{fig1} and \ref{fig2} we present mid-infrared images and
integrated spectra of the galaxies in our sample. All galaxies show
evidence of star formation activity as it's indicated by the presence
of the Unidentified Infrared Bands (UIBs) in their spectra
(\cite{leger}). Two trends became apparent from those figures:

As we move from early stage interactions to mergers the continuum at
12-16\,$\mu$m is rising very steeply. This continuum is attributed
to Very Small Grains (VSGs) with radius less than 10\,nm
(\cite{desert}), and is prominent in regions actively forming stars.
It reaches its peak in NGC6240 and Arp220, which host massive
starbursts, and progressively becomes flatter in post-starburst
systems.

The fraction of the mid-infrared flux associated with the UIB features
decreases when we reach the starburst face. This can be easily seen by
observing the strength of the 7.7$\mu$m feature. This could be due to
the fact than in massive starbursts one has numerous young stars and
their associated HII regions. As a result the filling factor of the
photodissociation regions where UIBs form would decrease as well as
the corresponding UIB emission.

\begin{figure}[h]
\resizebox{\hsize}{!}{\includegraphics{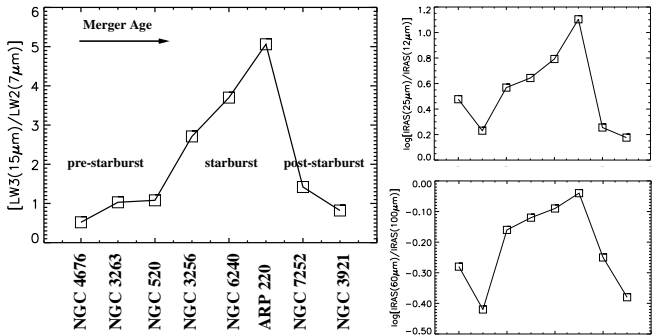}}
\caption{A comparison of the variation of the ISOCAM LW3/LW2 flux ratio
along the merging sequence, with the well known IRAS flux ratios. Note
how well the ISOCAM starburst diagnostic follows the evolution of the
star forming activity/merger age of the sequence. One may effectively
consider the LW3/LW2 ratio as a tracer of the location of FIR peak of
the bolometric luminosity. The IRAS 12$\mu m$ and 25$\mu m$ fluxes,
corrected for the extent of the galaxies, have been kindly provided by
D.B. Sanders (Univ.  Hawaii).
\label{fig3}}
\end{figure}

One can attempt to quantify this phenomenon by observing the variation
of the global mid-infrared colors of the galaxies. We present the flux
ratio the total broad band LW3(12-18\,$\mu m$)/LW2(5-8.5\,$\mu$m) for
our galaxies in Figure 3. This has been proposed as an indicator of
the fraction of the VSG continuum to the UIB feature emission and it
is close to 1 for quiescent star formation (\cite{boselli}). As we
clearly see this ratio presents a monotonic variation with the
intensity of the star formation activity in the galaxies.

We can also examine how the IRAS colors vary across the same sequence
of galaxies. Of particular interest is the IRAS60$\mu m$/IRAS100$\mu
m$ ratio since this indicates the location of the peak of the spectral
energy distribution. The correlation of the ISOCAM LW3/LW2 diagnostic
ratio with the IRAS colors is apparent. The only discrepant point is
NGC4676, but this can be understood since the one of the galaxies has
an old stellar population which can contribute to the mid-infrared
emission (\cite{hibbard}).

This result suggests that {\it even though the bolometric luminosity of
interacting luminous infrared galaxies is found at $\lambda \geq 40
\mu m$, the study of the mid-infrared spectral energy distribution can be
used to trace the location of the far-infrared peak}. Moreover, the
higher spatial resolution one can achieve in the mid-infrared
(\cite{keck}) would further facilitate the identification of the most
active regions in the galaxies (ie. \cite{felix})

\subsection{The SIRTF/IRS contribution}

Strong absorption by dust can distort the apparent morphology of
interacting galaxies, hiding the main heating source and revealing to
us only reprocessed radiation. Consequently, ISO estimates of the
absorption using measurements of line strengths (\cite{lutz}) may be
biased towards lower limits.  The use of the depth of the 9.6$\mu$m
silicate absorption feature could be an alternative, but as seen in
Arp220 this feature can often be nearly saturated and the underlying
continuum is poorly determined.

IRS, the infrared spectrograph (\cite{houck}) on board SIRTF, with a
spectral coverage from 5.3 to 40 $\mu$m will enable to address this
issue. Improved estimates on the absorption could be obtained using
the depth of both silicate absorption bands (at 9.6 and 18 $\mu$m) and
the shape of the mid-infrared continuum (\cite{dudley}).  Moreover,
the superb sensitivity and good spatial and spectral resolution of IRS
will allow us to further expand the diagnostic of Fig.3 in fainter
more distant systems.

\appendix

\end{document}